\documentclass[sn-mathphys-num]{sn-jnl}

\usepackage[utf8]{inputenc}
\usepackage{graphicx}
\usepackage{dcolumn}
\usepackage{hyperref}
\hypersetup{
  colorlinks=true,        
  linkcolor=blue,         
  citecolor=blue,         
  urlcolor=blue           
}
\usepackage{amsmath}
\usepackage{amssymb}
\usepackage{cancel}
\usepackage{bm}
\usepackage{booktabs}
\usepackage{mathtools}
\usepackage{times}
\usepackage[]{lineno}

\usepackage[caption=false]{subfig}

\newcommand{\Mtov}{M_\mathrm{TOV}}
\newcommand{\Ctov}{C_\mathrm{TOV}}
\newcommand{\ntov}{n_\mathrm{TOV}}
\newcommand{\Ptov}{p_\mathrm{TOV}}
\newcommand{\Rtov}{R_\mathrm{TOV}}
\newcommand{\msol}{M_\odot}
\newcommand{\Rm}{R_{1.4}}

\renewcommand{\hat}{\widehat}
\newcommand{\eg}{e.g.,~}
\newcommand{\ie}{i.e.,~}

\newcommand{\revtext}[1]{\textcolor{black}{#1}}
\newcommand{\revtextB}[1]{\textcolor{black}{#1}}

\makeatletter
\g@addto@macro\bfseries{\boldmath}
\makeatother

\begin{document}

\title{\textbf{Constraining the equation of state in neutron-star cores via the long-ringdown signal}}

\author[1]{\fnm{Christian} \sur{Ecker}}

\author[1,2,3]{\fnm{Tyler} \sur{Gorda}}

\author[4]{\fnm{Aleksi} \sur{Kurkela}}

\author[1,5,6]{\fnm{Luciano} \sur{Rezzolla}}

\affil[1]{\orgdiv{Institut f\"ur Theoretische Physik}, \orgname{Goethe
    Universit\"at}, \orgaddress{\postcode{60438}, \city{Frankfurt am
      Main}, \country{Germany}}}

\affil[2]{\orgdiv{ExtreMe Matter Institute EMMI}, \orgname{GSI
    Helmholtzzentrum f\"ur Schwerionenforschung GmbH}, \postcode{64291},
  \orgaddress{Darmstadt}, \country{Germany}}

\affil[3]{\orgdiv{Department of Physics}, \orgname{Technische
    Universit\"at Darmstadt}, \postcode{64289} \orgaddress{Darmstadt},
  \country{Germany}}

\affil[4]{\orgdiv{Faculty of Science and Technology}, \orgname{University
    of Stavanger}, \orgaddress{\city{Stavanger}, \postcode{4036},
    \city{Stavanger}, \country{Norway}}}

\affil[5]{\orgname{Frankfurt Institute for Advanced Studies},
  \orgaddress{\postcode{60438},
    \city{Frankfurt}, \country{Germany}}}

\affil[6]{\orgdiv{School of Mathematics} \orgname{Trinity College},
  \orgaddress{\city{Dublin} \postcode{2}, \country{Ireland}}}

\vspace{2cm}

\abstract{Multimessenger signals from binary neutron star (BNS) mergers
  are promising tools to infer the largely unknown properties of nuclear
  matter at densities that are presently inaccessible to laboratory
  experiments. The gravitational waves (GWs) emitted by BNS merger
  remnants, in particular, have the potential of setting tight
  constraints on the neutron-star equation of state (EOS) that would
  complement those coming from the late inspiral, direct mass-radius
  measurements, or ab-initio dense-matter calculations. To explore this
  possibility, we perform a representative series of general-relativistic
  simulations of BNS systems with EOSs carefully constructed so as to
  cover comprehensively the high-density regime of the EOS space. From
  these simulations, we identify a novel and tight correlation between
  the ratio of the energy and angular-momentum losses in the late-time
  portion of the post-merger signal, \ie the ``long ringdown'', and the
  properties of the EOS at the highest pressures and densities in
  neutron-star cores. When applying this correlation to post-merger GW
  signals, we find a significant reduction of the EOS uncertainty at
  densities several times the nuclear saturation density, where no direct
  constraints are currently available. Hence, the long ringdown has the
  potential of providing new and stringent constraints on the state of
  matter in neutron stars in general and, in particular, in their cores.}

\maketitle

The densest matter in the observable universe is found in the cores of
neutron stars (NSs), where gravity compresses it to supernuclear
densities, exceeding manyfold the nuclear density of $n_{\rm sat} = 0.16$
baryons/fm$^3$. While the behaviour of pressure and density in such
cores, that is, the equation of state (EOS) of strongly interacting
matter, remains an open question, a precise determination of the EOS of
NSs would provide precious insights on the phase diagram of Quantum
Chromodynamics (QCD).

In recent years, there have been remarkable advances in the inference of
the EOS thanks to rapidly advancing observations of NSs and theoretical
ab-initio calculations (see, \eg~\cite{Annala2023}). In addition, the
observation of the GW signal from the late inspiral of the BNS merger
event GW170817 demonstrated the potential of GW measurements to constrain
the EOS by setting limits on the tidal deformabilities of the
inspiralling NSs, which are tightly correlated with the EOS at densities
$\sim 3\, n_\mathrm{sat}$ and reached by NSs prior to merger~(see, \eg
Refs.~\cite{Baiotti2016, Paschalidis2016, Radice2020b} for
reviews). Finally, the analysis of the electromagnetic counterpart
associated with GW170817 provided convincing evidence for the formation
of a hypermassive neutron star (HMNS) that collapsed into a black hole
\revtext{over a timescale that, under a number of assumptions, has been
  estimated to be of approximately one second after the
  merger~\cite{Gill2019, Murguia-Berthier2020}}. Because HMNSs are
expected to reach densities that are significantly higher than those in
the stars before the merger, the study of their properties opens up the
opportunity to directly determine the EOS up to the highest densities in
the observable Universe.

This potential will be fully realized with the upcoming third-generation
GW observatories~\cite{Punturo:2010, Evans2021}, whose high sensitivity
at frequencies larger than 1 kHz allows them to detect with high
signal-to-noise ratios (SNR) also the post-merger signal from the
HMNS~\cite{Baiotti2016, Paschalidis2016, Radice2020b}. This has important
consequences, as a number of studies have shown that the most prominent
features of the \revtextB{power spectral density} (PSD) of the post-merger signal,
\ie the $f_1-f_3$ ``peaks'' of the spectrum, correlate with the
underlying EOS models~(see, \eg~\cite{Bauswein2015, Takami2015,
  Rezzolla2016, DePietri2018, Kiuchi2022, Breschi2022a}). We here propose
an improvement on this approach that concentrates on a specific and late
part of the post-merger signal and that allows for a direct determination
of the EOS at the highest densities.

More specifically, just like the ringdown of a perturbed black hole
contains precise information on the black-hole properties, we show that
the late-time, attenuated GW signal produced by the HMNSs between $\sim
1$ and $\sim 15\,{\rm ms}$ holds a similar potential in showing a clear
correlation with the maximum densities and pressures of the EOS. We refer
to this late-time signal as to the ``\emph{long ringdown}'' since its
characteristic damping time is much longer than the typical damping time
of a black hole with similar mass. The origin of this long ringdown is to
be found in the fact that $\sim 10\,{\rm ms}$ after the merger -- when
the GW amplitude is still comparatively large -- the HMNS exhibits a
quasi-stationary dynamics, with a mostly axisymmetric equilibrium and
small $\ell=2, m=2$ deformations that are responsible for an almost
constant-frequency, constant-amplitude GW emission. Under these
conditions, a simple toy model, as that introduced in
Ref.~\cite{Takami2015}, is sufficient to show that thanks to the
equilibrium achieved during this stage, the radiated GW energy and
angular momentum follow a linear relation and correlate strongly with the
EOS at high densities. Hence, the observation of the long ringdown at a
large SNR represents a novel and faithful probe of the largest densities
and pressures of the remnant's EOS.

\begin{figure}
  \centering
  \includegraphics[width=\textwidth]{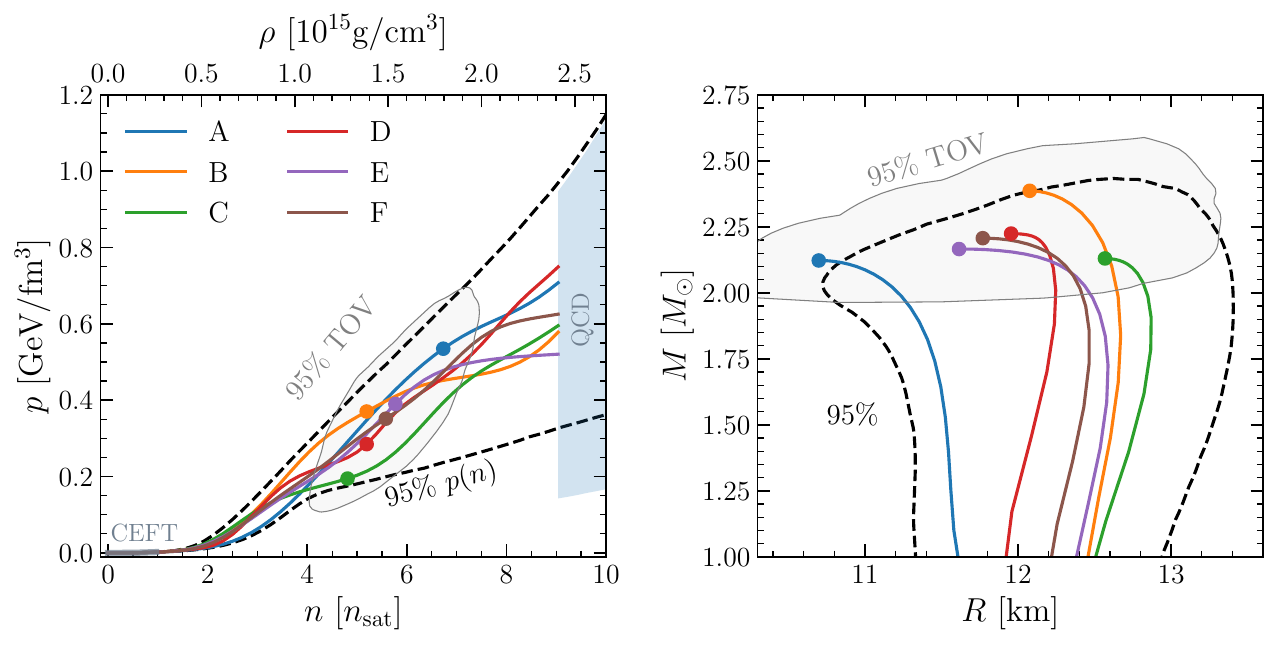}
  \caption{\textbf{The ``golden'' EOSs.} \textit{Left panel:} Solid lines
    of different colors show the six golden EOSs (${\rm A}$--${\rm F}$)
    in the $(p,n)$ plane. The dashed black lines show the $95\%$ credible
    intervals of all possible EOSs, while the CEFT and QCD bounds are
    shown with shaded areas (gray and light blue, respectively). Colored
    filled circles show the TOV points of the golden EOSs, while the
    solid light gray line is the $95\%$ credible interval for all TOVs.
    \textit{Right panel:} The same as in the left but shown in the
    $(M,R)$ plane.}
\label{fig:fig1}
\end{figure}

In order to quantify this novel correlation, we perform a suite of
general-relativistic simulations of BNS mergers with EOSs carefully
constructed so as to comprehensively cover the currently allowed space of
parameters. More precisely, we employ a large posterior sample of
model-agnostic, Gaussian-process (GP) based, zero-temperature EOSs of NS
matter from~\cite{Gorda:2022jvk} that is conditioned with constraints from
the tidal-deformability measurement of GW170817, radio measurements of
high-mass pulsars, combined mass-radius measurements from X-ray
pulse-profile modeling of isolated NSs, as well as with low-energy
nuclear-theory constraints from chiral effective field theory (CEFT) and
high-energy particle-theory bounds from perturbative QCD. The significant
breadth of EOSs in the ensemble reflects the current level of uncertainty
in the determination of the EOS. Because it is computationally
prohibitively expensive to scan a large number of EOSs, we reduce the
full ensemble to a smaller sample of six ``golden'' EOSs that maximizes
the variation in the following four NS parameters: the maximum (TOV) mass
of an isolated, nonrotating NS $\Mtov$, its \revtext{dimensionless}
compactness $\Ctov:=\Mtov/\Rtov$, where $\Rtov$ is the corresponding
radius, the central pressure $\Ptov$, and the radius of a typical
$1.4\,\msol$ NS $\Rm$. By performing a principal-component analysis [see
  supplemental material (SM) for details], we select six EOSs that are
located in the center (EOS labeled ${\rm F}$) and distributed on the
boundary (EOSs labelled ${\rm A-E}$) of the $68\%$-credible region in the
four-dimensional space spanned by the NS parameters (see
Fig.~\ref{fig:PCA} of SM). \revtext{We have chosen this region so that
  our sample characterises the distribution where most of the posterior
  weight is; a different choice of, \eg $95\%$ would consist of EOSs that
  are already in tension with observations and would not necessarily
  characterise the distribution as faithfully as the sample would be
  sensitive to the tails of the distribution.}

These six EOSs are shown in Fig.~\ref{fig:fig1} in the pressure--number
density $(p,n)$ plane (left panel), along with the corresponding
mass-radius relationships for nonrotating stars (right panel). We note
that, by construction, our global sample of EOSs, and hence also the
golden EOSs, do not contain strong phase transitions, which could lead to
a larger EOS space consistent with astrophysical bounds
\cite{Gorda:2022b}. This choice allows us to focus our attention on
smooth EOSs and to build an understanding of their phenomenology, leaving
the exploration of EOSs with phase transitions to a subsequent work where
we will employ the approach in~\cite{Annala2019,Altiparmak:2022}.

While NSs during the inspiral stage can be described also when neglecting 
the temperature dependence of the EOS, during and after the merger 
shock-heating effects lead to non-negligible temperatures inside the 
merger remnant. To model and
approximate these heating effects, the zero-temperature EOSs selected in
our sample are modified in what is normally referred to as a ``hybrid
EOS'', where the cold part of the EOS is combined with an ideal-gas EOS,
thus providing an effective-temperature contribution. While this is an
approximation, it does not affect the properties of the correlation and
we have adopted an adiabatic-index value of $\Gamma_\mathrm{th} = 1.75$,
\revtext{which is close to the optimal value ($\Gamma_\mathrm{th}\approx
  1.7$) suggested in~\cite{Figura2020} and} on average mimics well the
temperature dependence of microscopic
constructions~\cite{Demircik:2021zll} (see SM for details). We have
verified that all of the qualitative properties of the GW emission from
the HMNS presented here are preserved when considering also
self-consistent temperature-dependent EOSs, such as the
V-QCD~\citep{Demircik:2021zll} or the HS-DD2 EOS~\cite{Hempel2010} (see
SM).

Using these six golden EOSs, we performed a series of
general-relativistic BNS merger simulations and extracted the emitted GW
signal starting from about $15\,{\rm ms}$ before merger until $30\,{\rm ms}$
after. From these simulations, we compute the instantaneous GW frequency
$f_\mathrm{GW}$, the radiated energy $E_\mathrm{GW}$, and angular
momentum $J_\mathrm{GW}$. The binaries have been constructed with
parameters that are consistent with those measured for GW170817, \ie with
fixed chirp mass $\mathcal{M}_{\rm chirp} = 1.18\,M_\odot$ and three
different ratios $q := M_2 / M_1 =0.7,0.85,1$ of the binary constituent
masses $M_1$ and $M_2$. From a qualitative point of view, the dynamics of
the six binaries reflects what has been found by a large number of works
(\eg~\cite{Baiotti08, Bauswein:2010dn, Takami2015, Kastaun2016,
  Hanauske2016, DePietri2018, Most2018b, Tootle2022}), with an HMNS
attaining a metastable equilibrium a few milliseconds after the merger
and then emitting GW radiation at frequency that is almost constant in
time and around the characteristic $f_2$ frequency of the post-merger
PSD~\cite{Rezzolla2016}. \revtext{Here, we instead focus on the rates at
  which energy and angular momentum are radiated by the HMNS when it has
  reached a quasi-stationary equilibrium at about $10\,{\rm ms}$ after
  the merger (see also~\cite{Bernuzzi2015b, Chaurasia2020}). }


\begin{figure}
  \centering
  \includegraphics[width=0.65\textwidth]{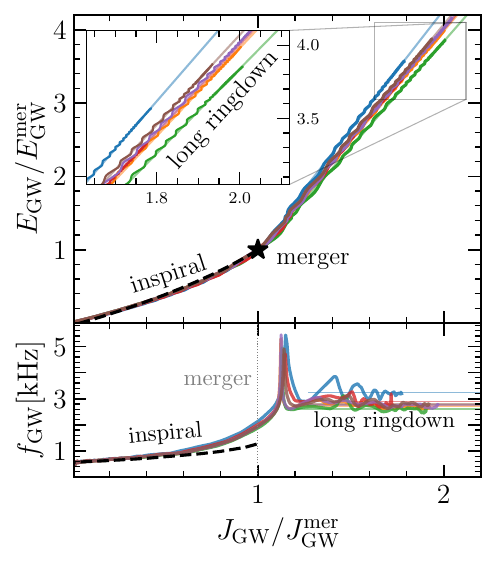}
  \caption{\textbf{Radiated GW energy and angular momentum.} \textit{Top
      panel:} Using the same color convention as in Fig.~\ref{fig:fig1},
    we report the energy $E_\mathrm{GW}$ and angular momentum
    $J_\mathrm{GW}$ emitted in GWs for the golden EOSs when normalized to
    the values at merger, indicated by a star. The evolution expected
    from the post-Newtonian inspiral is shown with a dashed black
    line. The inset offers a magnification of the long ringdown with the
    corresponding linear slopes indicated by thin lines of the same
    color. \textit{Lower panel:} As in the top panel but in terms of the
    instantaneous GW frequency $f_\mathrm{GW}$. The data refers to
    equal-mass binaries but very similar behaviour is found also for
    unequal-mass binaries (see Fig.~\ref{fig:fig2_qneq1} in the SM for
    details).}
\label{fig:fig2}
\end{figure}

Figure~\ref{fig:fig2} displays the most salient results of the six
equal-mass binaries by showing in the top part the evolution of the
radiated GW energy $E_\mathrm{GW}$ and angular momentum $J_\mathrm{GW}$
normalized by their values from the start of the simulations till the
merger\footnote{The specific starting time of the simulation, or the time
when the signal enters the detector's sensitivity curve, are not relevant
here, since the energy and angular momentum radiated in the entire
inspiral phase are negligibly small compared to the losses during and
after the merger~\cite{Zappa2018, Nathanail2021}.},
$E^\mathrm{mer}_\mathrm{GW}$ and $J^\mathrm{mer}_\mathrm{GW}$, where, as
customary, $t_{\rm mer}$ is defined as the moment of maximum-amplitude
strain. Note the significant change in the evolution between the inspiral
phase and the post-merger. In the former, all binaries -- indicated with
the same color convention as in Fig.~\ref{fig:fig1} -- follow essentially
the same trajectory in the $(E_\mathrm{GW}, J_\mathrm{GW})$ plane, which
is well captured by the post-Newtonian (PN) approximation shown as a
black dashed line (we use a reference tidal deformability
$\tilde\Lambda=580$ in the $5$PN-order Taylor-T2 model of the PyCBC
library~\cite{Biwer:2018}). Obviously, there are differences in evolution
of the six binaries that are generated by the different tidal
deformabilities, but these differences are minute when compared with
those that emerge after the merger, when the different evolutions become
visibly distinct. More importantly, it is remarkable that in the latter
part of the signal, \ie in the long ringdown, the normalized radiative
losses in $E_\mathrm{GW}$ and $J_\mathrm{GW}$ are linearly related, as
\revtext{first noted in \cite{Bernuzzi2015b, Chaurasia2020} and}
clearly shown in the inset reporting a magnification of the long
ringdown. This apparently striking behaviour has a rather simple
explanation in terms of the Newtonian quadrupole formula applied to a
rotating system with an $\ell=2=m$ deformation, as in the toy model of
Ref.~\cite{Takami2015}. In this case, one can show the identity
$\dot{E}_\mathrm{GW} / \dot{J}_\mathrm{GW} = d{E}_\mathrm{GW} /
d{J}_\mathrm{GW} = f_\mathrm{GW}/(2\pi)$, where we use a dot to indicate
a time derivative (see SM). Stated differently, during the long ringdown,
the HMNS behaves essentially as a rotating $m=2$ deformed quadrupole and
radiates GW energy and angular momentum that are linearly
related. Indeed, we have verified that more than $97\%$ of the
gravitational wave amplitude arises from this dominant mode. We note
that while Fig.~\ref{fig:fig2} refers to equal-mass binaries, a perfectly
analogous behaviour is also realized by unequal-mass binaries, which we
do not show here for clarity, but that can be found in the SM (see
Fig.~\ref{fig:fig2_qneq1}).

The importance of the results summarised in Fig.~\ref{fig:fig2} is that
binaries whose HMNS signal can be measured with high SNR, and hence for
which \revtext{the radiated GW energy and angular momentum} can be
measured more accurately, offer a key to access the properties of the EOS
at the highest densities and pressures, \ie $\ntov$ and $\Ptov$. Before
discussing how this can be done and to what precision, we need to make a
few important remarks. First, the physical picture presented in
Fig.~\ref{fig:fig2} in terms of the slope between ${E}_\mathrm{GW}$ and
${J}_\mathrm{GW}$ can also be drawn in terms of the instantaneous GW
frequency $f_\mathrm{GW}$, which is instead presented in the bottom part
of Fig.~\ref{fig:fig2}. This panel shows in fact that during the long
ringdown $f_\mathrm{GW}$ also asymptotes to an essentially constant
value, $f_\mathrm{GW} = {f}_\mathrm{rd}\simeq {\rm const}.$ \revtext{(see
  also~\cite{Bernuzzi2015b} where this behaviour was first mentioned)},
and---\revtext{assuming that the signal is dominated by the $\ell=2=m$
  mode}---this is the same value that can be deduced from the slope
between ${E}_\mathrm{GW}$ and ${J}_\mathrm{GW}$. Hence, while measuring
${f}_\mathrm{rd}$ in the long ringdown is conceptually analogous to
measuring the slope, we have found that the latter is more robust as it
is easier to fit an approximately linear function, \ie ${E}_\mathrm{GW}$
vs ${J}_\mathrm{GW}$, than the average of an oscillating and potentially
noisy function, \ie ${f}_\mathrm{GW}$;\footnote{\revtext{This is because
  the variance of the slope in a linear fit of data $\{(x_i, y_i)\}$ is
  suppressed by an additional factor ${\sum_i(x_i - \langle x
    \rangle)^2}$ when compared to the residual variance of the fit.}}
the extrapolated slopes are shown with \revtext{thin} lines of the
corresponding color in Fig.~\ref{fig:fig2}. Second, the long-ringdown
frequency ${f}_\mathrm{rd}$ is close to but different from the main-peak
frequency in the post-merger PSD, \ie $f_2$, which is traditionally
advocated as a good proxy for the EOS~\cite{Bauswein2015, Takami2015,
  Rezzolla2016, DePietri2018, Kiuchi2022, Breschi2022}. This is because
${f}_\mathrm{GW}$ oscillates wildly right after the merger and hence
$f_2$ collects power from frequencies that are both larger and smaller
than ${f}_\mathrm{rd}$ (see lower part of Fig.~\ref{fig:fig2}), thus
increasing the uncertainty in its measurement. Stated differently, the
difference between $f_2$ and $f_{\rm rd}$ is that the former collects
power over a very broad window in time starting from the merger, while
the latter contains information during a very narrow window around the
long ringdown. Finally, our analysis reveals that the correlations
between the GW signatures (\ie $f_{2}$ and $f_{\rm rd}$) and the
properties of the EOS (\ie $\ntov$ and $\Ptov$) are statistically
different. \revtext{In particular, we have measured the
  Pearson-correlation coefficients $r(X, Y) := \operatorname{cov}(X,Y) /
  (\sigma_X \sigma_Y)$ between the data from our golden EOSs to be
  $r(dE_{\rm GW}/dJ_{\rm GW}, \Ptov)=0.877$ and $r(dE_{\rm GW}/dJ_{\rm
    GW}, \ntov)=0.917$ in the case of the slope, and $r(f_2,
  \Ptov)=0.792$ and $r(f_2, \ntov)=0.865$ in the case of $f_2$, thus
  indicating that there is a strong correlation in both cases, but also
  that this is stronger for the long-ringdown frequency.}

\begin{figure}
  \centering
  \includegraphics[width=\textwidth]{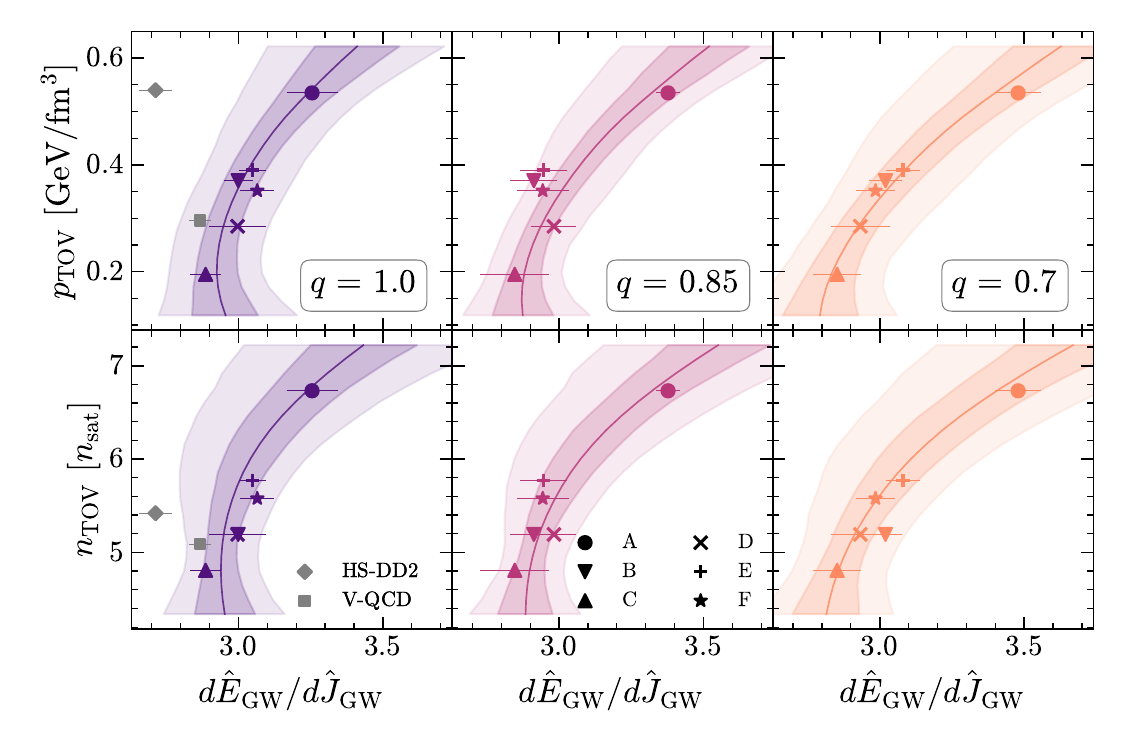}
  \caption{\textbf{Correlation between the long-ringdown slope and the
      TOV properties.} Data and fits illustrating the correlations
    between the slope of the radiated quantities $\hat{E}_\mathrm{GW}$
    and $\hat{J}_\mathrm{GW}$ normalized to the merger values, and
    $p_\mathrm{TOV}$ (top row) or $n_\mathrm{TOV}$ (bottom row), and for
    different mass ratios (different columns). The dark (light) shaded
    regions denote $68\% (95\%)$ credible intervals for the bilinear
    model~\eqref{eq:bilinear_model}, while the solid lines denote the
    mean value. A marginalization over the remaining EOS quantity in the
    bilinear model was performed. \revtext{Also shown are data points for two
      microscopic EOS models, one of which (HS-DD2) is disfavored by
      astrophysical data.}}
\label{fig:fig3}
\end{figure}

To illustrate how to make use of the long ringdown to set constraints on
the EOS at the highest densities and pressures, Fig.~\ref{fig:fig3} shows
the tight correlation between the normalized slope during the long
ringdown $d\hat{E}_\mathrm{GW} / d\hat{J}_\mathrm{GW}$, where
$\hat{E}_\mathrm{GW} := E_\mathrm{GW} / E^\mathrm{mer}_\mathrm{GW}$ and
$\hat{J}_\mathrm{GW} := J_\mathrm{GW} / J^\mathrm{mer}_\mathrm{GW}$, and
the highest pressure $p_{\rm TOV}$ and density $n_{\rm TOV}$ reached in
nonrotating NSs. The six different panels of Fig.~\ref{fig:fig3} are
organized so as to show in the three columns the correlations for the
three different mass ratios considered ($q=1,0.85$ and $0.70$ from left
to right) and in the two rows the variation with the maximum pressure
(top row) and the maximum density (bottom row). \revtext{It is
  straightforward to appreciate from the six panels that the correlation
  is strong and we find it quite striking that measuring the
  long-ringdown slope of a low-mass NS can provide precise information on
  the properties of matter at the highest densities and pressures
  realized in nature and which are well above those probed in the merger
  remnant.}

To quantify the strength of the correlation, we consider a bilinear model
fit to the data from our simulations
\begin{align}
  \frac{d\hat{E}_\mathrm{GW}}{d\hat{J}_\mathrm{GW}} = \beta_0 + \beta_1\,
  p_\mathrm{TOV} + \beta_2\, n_\mathrm{TOV} + \beta_3\, q + \beta_4 \, q \,
  p_\mathrm{TOV} + \beta_5\, q \, n_\mathrm{TOV} + \beta_6 \,
  p_\mathrm{TOV} \, n_\mathrm{TOV}\,.
  \label{eq:bilinear_model}
\end{align}
Note that the ansatz \eqref{eq:bilinear_model} ignores quadratic terms in
$q, p_\mathrm{TOV}$, and $n_\mathrm{TOV}$ as these provide only marginal
improvements to the fit (for $q$) or break it (for $p_\mathrm{TOV},
n_\mathrm{TOV}$). After fitting this model over the time window $t -
t_\mathrm{mer} \in [1, 30]\,{\rm ms}$, we obtain a probability
distribution for the model parameters $\beta_i$ and, in turn, for the
long-ringdown slope given the parameters of the EOS $P(d
\hat{E}_\mathrm{GW} / d \hat{J}_\mathrm{GW} | p_\mathrm{TOV},
n_\mathrm{TOV}, q)$, which we use to produce the $68\%$ ($95\%$) credible
intervals for $d\hat{E}_\mathrm{GW} / d \hat{J}_\mathrm{GW}$ shown in
dark (light) shading in each of the panels in Fig.~\ref{fig:fig3}. These
intervals are produced by marginalizing over the other EOS variables
using the underlying probability distribution $P(p_\mathrm{TOV},
n_\mathrm{TOV})$ from the EOS ensemble.

Clearly, the bilinear model~\eqref{eq:bilinear_model} reproduces well
long-ringdown slopes, where the distribution of model parameters
$\bm{\beta} := (\beta_0, \ldots, \beta_6)$ are given by a multivariate
Gaussian distribution with a mean $\bar{\bm{\beta}} = ( 1.78,  0.72,
-1.44,  1.90,  -1.74, -1.14,  3.61)$ and a covariance matrix
$\operatorname{cov}(\bm{\beta})$ reported in Table~\ref{tab1} of the SM,
with $p_\mathrm{TOV}$ and $n_\mathrm{TOV}$ expressed in units of ${\rm
GeV/fm}^{3}$ and ${\rm fm}^{-3}$, respectively.

\begin{figure}
  \centering
  \includegraphics[width=\textwidth]{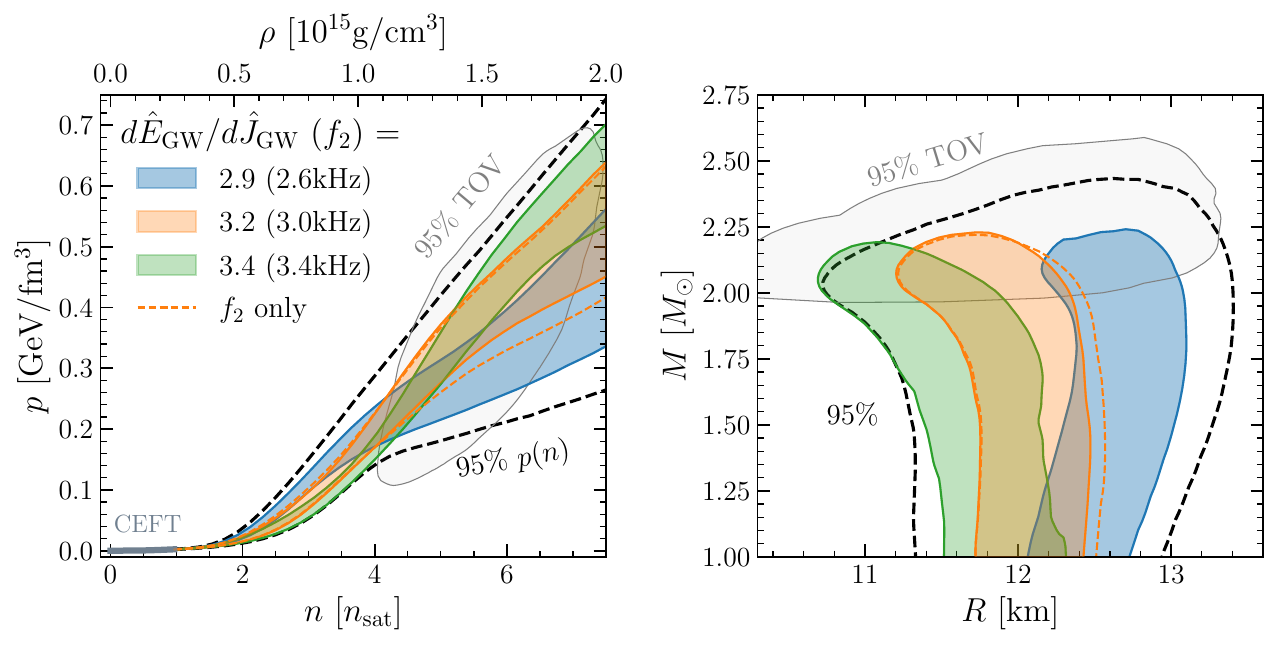}
  \caption{\textbf{Impact of slope measurements on the EOS and NS
      properties.}  \textit{Left panel} Using the same convention as in
    Fig.~\ref{fig:fig1}, we show the $68\%$ credibility values on the
    $(p,n)$ plane from potential \textit{joint} measurements of the
    long-ringdown slope and $f_2$ assuming a flat distribution for
    $q\in[0.7,1]$ and an uncertainty of $\pm 4\%$ on the slope and $\pm
	4\%$ on $f_2$. Shown with a dashed line is the result for using only
    the $f_2$ measurement. \textit{Right panel} the same but in the
    $(M,R)$ plane.}
\label{fig:pn_mr_measured}
\end{figure}

\revtext{Having pointed out a novel correlation between the properties of the
  long-term GW signal and the properties of the EOS at the
  \textit{highest density}, we now take our analysis a step further and
  show how our results, in conjunction with a future post-merger GW
  detection, can be used to constrain the EOS at \textit{all densities}.}
More specifically, to demonstrate the effect of a future measurement of
$d \hat{E}_\mathrm{GW} / d \hat{J}_\mathrm{GW}$, we use Bayes's theorem
to infer the combined EOS and NS properties
\begin{equation}
    P(\mathrm{EOS}, \mathrm{NSs} | \mathrm{data}) = \frac{P(\mathrm{data} |
      \mathrm{EOS}, \mathrm{NSs}) P(\mathrm{EOS}, \mathrm{NSs})}{P(\mathrm{data})}\,.
\end{equation}
In particular, we can use as an additional piece of the likelihood function
$P(\mathrm{data} | \mathrm{EOS}, \mathrm{NSs})$ the integral of our bilinear
model over the likelihood of the measurement (see SM for the details). Assuming
a Gaussian measurement of the long-ringdown slope, as well as a uniform
distribution on the mass ratio $q \in [0.7, 1.0]$ from the measurement, we
display in Fig.~\ref{fig:pn_mr_measured} the resulting constraints at $68\%$
credibility.

The left panel of Fig.~\ref{fig:pn_mr_measured} considers three values of
the measured long-ringdown slope, \ie $d\hat{E}_\mathrm{GW} /
d\hat{J}_\mathrm{GW} = 2.9, 3.2$ and $3.4$ with an error estimate of
$8\%$, \textit{joint} with the measurements of the frequency $f_2=2.6,
3.0, 3.4\,{\rm kHz}$ with an error of $8\%$~\cite{Bose2017,
  Breschi2022}\footnote{\revtext{Considering that the standard deviation
  of the measurement of $d\hat{E}_\mathrm{GW} / d\hat{J}_\mathrm{GW}$ has
  been found to be of about $3\%$, our error estimates in
  Fig.~\ref{fig:pn_mr_measured} have been rather conservative. Only when
  using a distinct and more extensive analysis taking into account
  realistic GW-signal-processing pipelines will it be possible to set
  less conservative error estimates.}}. \revtext{For this analysis, we
  use a second bilinear model to fit the $f_2$ data, whose quality is
  found to be as good as the above fit to $d \hat{E}_\mathrm{GW} / d
  \hat{J}_\mathrm{GW}$.} Using different colors for the different
  measurements, it is then apparent that smaller (larger) values of the
  slope would constrain the EOS to have higher (lower) pressures at 2--4
  times nuclear saturation density while also having a smaller (larger)
  pressure and density for the maximally massive NSs. In turn, this leads
  to larger (smaller) radii for the most massive stars, as shown in the
  right panel of Fig.~\ref{fig:pn_mr_measured}. Note also how the
  measurement of the long ringdown provides an improvement of the
  corresponding posteriors obtained when measuring the $f_2$ frequency
  only (see dashed orange lines for the representative case of
  $d\hat{E}_\mathrm{GW} / d\hat{J}_\mathrm{GW} = 3.2$, $f_2=3.2\,{\rm
  kHz}$). \revtext{We should note that similarly large confidence
  intervals appear if we were to consider information on the slope only.
  Hence, Fig.~\ref{fig:pn_mr_measured} clearly highlights how the
  combination of information on the slope and on the $f_2$ frequency
  yields an increased accuracy in the properties of the EOS.}

Our concluding remarks are about the robustness of the new
correlation. We have already commented that the results apply
qualitatively unchanged when considering unequal-mass binaries or EOSs
with a consistent temperature dependence (see SM for details). In
addition, we have also verified that the same is true when considering
different values for the chirp mass. More specifically, taking
$\mathcal{M}_{\rm chirp}=1.13,1.22\,M_\odot$ instead of our fiducial
value of $1.18$ leads to differences in the slope that are significantly
smaller than those introduced by considering different EOSs (see SM for
details). Finally, and importantly, the long-ringdown slope is
essentially insensitive to different choices in the adiabatic index
$\Gamma_{\rm th}$, as we have verified by replacing our fiducial value of
$1.75$ with $\Gamma_{\rm th}=1.5$ or $2$ (see SM for details).

The preliminary study carried out here can be improved in a number of
ways, \eg by estimating the impact that large spins, strong magnetic
fields, neutrino emission, \revtext{strong first-order phase
  transitions,} and temperature-dependent EOSs \revtextB{and more generic treatments of the crust and sub-saturation density matter} have on the long-ringdown
slope. \revtextB{Additionally, the set of neutron-star parameters used in the principal-component analysis could be extended and optimized.  }
However, already now our new correlation between the radiated
energy and angular momentum during the long ringdown has the realistic
potential of significantly reducing the EOS uncertainty at the highest
densities realized in NSs for which no alternative observational
constraints are available to date. This potential may already be
exploited by the ongoing and near-future observations by the
LIGO-Virgo-Kagra collaboration, but it will surely play a fundamental
role in third-generation GW detectors, where the combined network of
Cosmic Explorer and Einstein Telescope are expected to detect $180$ BNS
signals per year with post-merger ${\rm SNR}>8$~\cite{Evans2021}. The
potential that these detectors have in measuring the long ringdown will
be explored in a forthcoming work.

\bmhead{Acknowledgements}

It is a pleasure to thank M. Cassing, M. Chabanov, E. Most, C. Musolino,
H. H.-Y. Ng, D. Radice, and K. Topolski for numerous discussions and
comments during the development of this work. Partial funding comes from
the Deutsche Forschungsgemeinschaft (DFG, German Research Foundation)
project-ID 279384907--SFB 1245, the State of Hesse within the Research
Cluster ELEMENTS (Project ID 500/10.006), by the ERC Advanced Grant
``JETSET: Launching, propagation and emission of relativistic jets from
binary mergers and across mass scales'' (Grant
No. 884631). C.~E. acknowledges support by the DFG through the CRC-TR 211
``Strong-interaction matter under extreme conditions''-- project number
315477589 -- TRR 211. L.~R. acknowledges the Walter Greiner Gesellschaft
zur F\"orderung der physikalischen Grundlagenforschung e.V. through the
Carl W. Fueck Laureatus Chair. The simulations were performed on the
local ITP Supercomputing Clusters Iboga and Calea and on HPE Apollo HAWK
at the High Performance Computing Center Stuttgart (HLRS) under the grant
BNSMIC.

\bmhead{Data Availability}

\revtextB{Data sets generated during the current study are available from the 
corresponding author on reasonable request.}

\section*{Supplemental Material}
\label{sec:methods}

In what follows we provide additional details on a number of aspects of
our analysis that we have omitted in the main text for compactness. These
refer to the approach followed for the selection of the golden EOSs, the
numerical techniques employed to simulate the binaries and extract the GW
signal, and a number of validations highlighting the robustness of the
correlation found between the EOS and the long-ringdown slope.

\subsection*{Selection of the golden EOSs}
For the agnostic construction of cold EOSs, we begin from the GP setup
presented in~\cite{Gorda:2022}, which we briefly review here. Below
densities of $n = 0.57\,n_\text{sat}$, we use the crust model by Baym,
Pethick, and Sutherland~\cite{Baym71b} 
\footnote{\revtextB{Note that the uncertainty associated with the 
crustal EOS has been recently discussed in \cite{Davis:2024nda}}.}.
Above this density, in the
interval $n = [0.57, 10] \,n_\text{sat}$, a GP regression is performed in
an auxiliary variable $\phi(n) := -\ln(1/c_s^2(n) - 1)$, where $c_s$ is
the sound speed, and where the prior for $\phi(n)$ is drawn from a
multivariate Gaussian distribution
\begin{equation}
\phi(n) \sim \mathcal{N}\bigl(-\ln (1/\bar{c}_s^2 - 1), K(n, n')\bigr)\,,
\end{equation}
with a Gaussian kernel $K(n, n') = \eta \exp \bigl( - (n - n')^2 / 2
\ell^2 \bigr)$~\cite{Gorda:2022}. The hyper-parameters $\eta$, $\ell$,
and $\bar{c}_s^2$ within these definitions are themselves drawn from
probability distributions
\begin{equation}
    \eta \sim \mathcal{N}(1.25, 0.25^2)\,,\quad \ell \sim
    \mathcal{N}(1.0\, n_\text{sat}, (0.2\, n_\text{sat})^2)\,,\quad
    \bar{c}_s^2 \sim \mathcal{N}(0.5, 0.25^2)\,.
\end{equation}
Below a density of $1.1\, n_\text{sat}$, the GP is conditioned with the
CEFT results from~\cite{Hebeler:2013nza}. In particular, the average
between the ``soft'' and ``stiff'' results from that work are taken as
the mean, while the difference between them is taken as the $90\%$
credible interval for the conditioning~\cite{Gorda:2022}. From this
GP, we draw sample of 120,000 EOSs. 

We impose the astrophysical observations referred to as ``Pulsars +
$\tilde\Lambda$'' in~\cite{Gorda:2022}. Explicitly, we use the
following three sets of observations:
\begin{enumerate}
\item heavy-pulsar mass constraints from radio astronomy. In particular,
  we use the constraints from PSR J0348$+$0432 with $M = 2.01 \pm 0.04\,
  M_\odot$~\cite{Antoniadis:2013pzd} and PSR J1624$-$2230 with $M = 1.928
  \pm 0.017\, M_\odot$~\cite{Fonseca2016}. We approximate the
  uncertainties from these measurements as normal distributions, which
  holds to good accuracy.
\item joint tidal-deformability and mass-ratio constraints from GW
  observations. We use the two-dimensional (2D) joint probability
  distribution for the tidal deformability and mass ratio using the
  \texttt{PhenomPNRT} waveform model with low-spin priors from Fig.~12
  of~\cite{LIGOScientific:2018hze}.
\item joint mass-radius measurements from X-ray pulse profile modeling.
  We use the 2D joint probability distribution for the mass and radius
  for PSR J0740$+$6620 using the NICER + XMM-Newton data from the right
  panel of Fig.~1 of~\cite{Miller:2021qha}.
\end{enumerate}
Within each of these measurements, we assume as our mass prior $P_0(M |
\text{EOS})$ a uniform distribution between $0.5\,M_\odot$ and
$M_\text{TOV}$.

Lastly, the GP is conditioned using information from high-density
perturbative QCD calculations, which are under theoretical control at
densities with baryon number chemical potential $\mu = 2.6\,{\rm GeV}$,
corresponding to $n \approx 40 n_\mathrm{sat}$. This information is
included in a conservative way, excluding those EOSs which cannot be
connected to the perturbative densities using any causal, mechanically
stable, and thermodynamically consistent interpolation in the density
interval $[10, 40] \, n_\mathrm{sat}$ \cite{Komoltsev:2021}. This is done
by conditioning the GP with the QCD likelihood function of
\cite{Gorda:2022}, where the uncertainty in the pQCD calculation at $\mu
= 2.6\,{\rm GeV}$ is taken into account by marginalizing over the
unphysical renormalization scale $X := 2 {\Sigma} / (3 \mu)$ in the range
$[0.5, 2]$, with $\Sigma$ the renormalization scale in the modified
minimal subtraction scheme.

We consider the posterior in the 4D space of $(M_\text{TOV},\,
C_\text{TOV}, \, \ln p_\text{TOV}, \, R_{1.4})$, within which we perform
a modified principal-component analysis to select a small sample of EOSs
that characterize the $68\%$ credible region of the distribution. This is
done as follows:
\begin{enumerate}
\item construct a normalized set of variables defined by $\hat{x} := (x -
  \mu_x) / \sigma_x$, with $\mu_x$, $\sigma_x$ the mean and standard
  deviation for the variable $x$.
\item construct the $4\times 4$ covariance matrix of these normalized
  variables.
\item calculate the eigenvalues $\lambda_i$ and eigenvectors $v_i$ of
  this covariance matrix, ordered by the magnitude on the eigenvalues.
\end{enumerate}
The orthogonal vectors $v_i$ define the principal components of the
distribution in the original 4D space, while the $\lambda_i$ characterize
the variance of the distribution in each of these directions. \revtext{In
  principle, one could generalize this analysis by considering additional
  uncorrelated variables to the four we have chosen, or even attempt to
  identify the optimal set of uncorrelated variables, but this is beyond
  the scope of this work.}

\begin{figure}
  \centering
  \includegraphics[width=\textwidth]{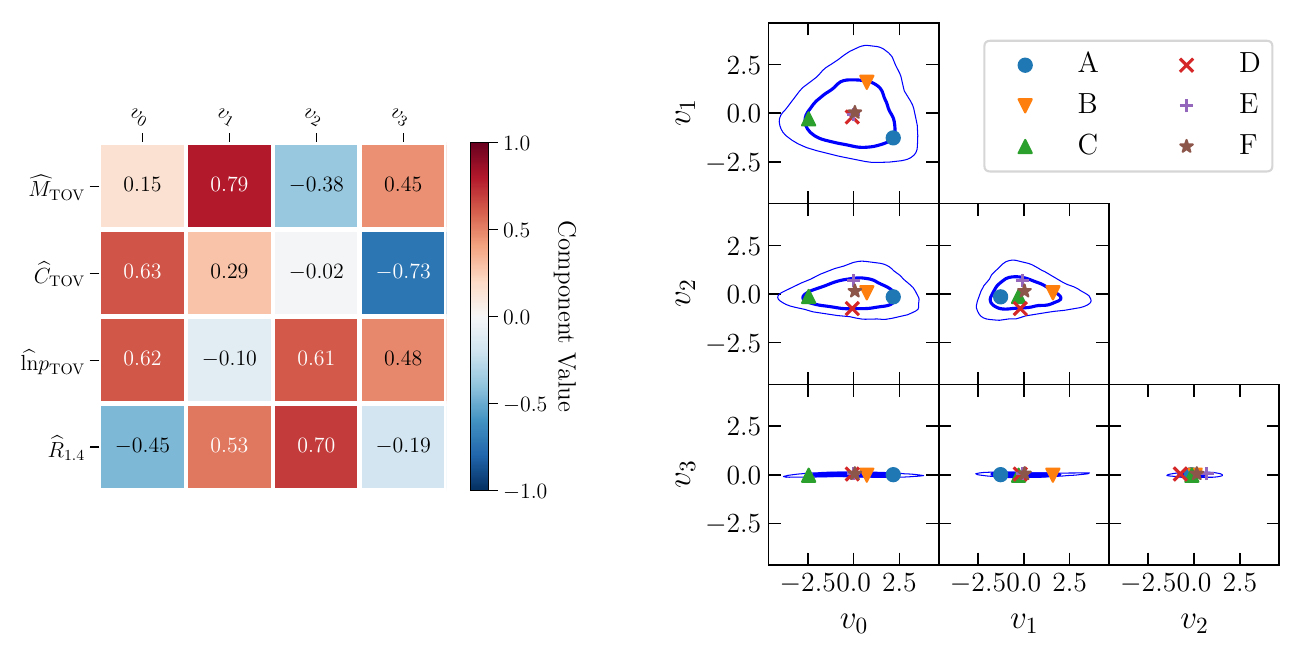}
    \caption{\textbf{Principal-Component Analysis.} \textit{Left panel:}
    Components of the principal-component vectors $v_i$ in terms of the original
    (normalized) coordinates $(\hat{M}_\mathrm{TOV}, \hat{C}_\mathrm{TOV},
    \hat{\ln}p_\mathrm{TOV}, \hat{R}_{1.4})$. \textit{Right panel:} Posterior
    distribution in the PCA coordinate system. Shown are $95\%$ (thin lines) and
    $68\%$ (thick lines) credible regions. Also shown are the six golden EOSs,
    with ${\rm A}$--${\rm E}$ lying on the $68\%$ contour by construction, and
    ${\rm F}$ lying at the center.}
\label{fig:PCA}
\end{figure}

Figure~\ref{fig:PCA} shows on the left the components of the $v_i$ in the
normalized coordinate system, while the right panel shows the posterior
distribution in the $v_i$ coordinate system. As seen in this figure, the
distribution is primarily 3D, with a prominent triangular component
within the plane spanned by the components $v_0$ and $v_1$. This
behaviour of the distribution clearly explains a well-known aspect to
anyone constructing agnostic models of EOSs, namely, that while it is
reasonable to model the variation in EOSs in terms of stiffness, \ie
$\hat{R}_{1.4}$, this choice does not cover all of the possible space of
parameters, which can be determined for instance in a principal-component
analysis. Finally, we select the six golden EOSs from our ensemble by
choosing EOSs that are near the extrema of the $68\%$ credible region and
one near the origin. With a standard principal-component analysis, these
points would be given by $\pm \sqrt{\lambda}_i v_i$ (no summation), which
we modify slightly to capture the triangular shape within the plane
spanned by $v_0$ and $v_1$. In this plane, we use the directions that
extremizes the $95\%$ credible regions. Having identified the relevant
points in the parameter space, we then select the golden EOSs
corresponding to one of these six points by finding the 30 closest EOSs
using the Euclidean metric in the full 4D space and selecting the one
with the highest posterior likelihood. We note that using the reduced 3D
metric obtained by dropping the $v_3$ component leads to the same final
golden EOSs. \revtextB{It is worth noting that using the corners of the
$68\%$ credibility contours for the golden EOS selection is a matter of
choice in how we represent the underlying distribution. The variability
of the simulations with the golden EOSs can be used as a proxy to
approximate the 68\%-credible regions that would be obtained if the full
GP ensemble was used. In our analysis, this choice of 68\% represents a
compromise between capturing the extrema of the EOS distribution and
assuring a sufficiently high posterior probability for the selected EOSs.
In particular, choosing instead the $95\%$ contour would represent the
same distribution with a selection of EOSs coming from the tails of the
EOS distribution, that is, EOSs that have significantly lower likelihood.
This choice does not affect the overall results to the extent that our
golden set characterises the features of the underlying distribution. }

\begin{figure}
  \centering
  \includegraphics[width=\textwidth]{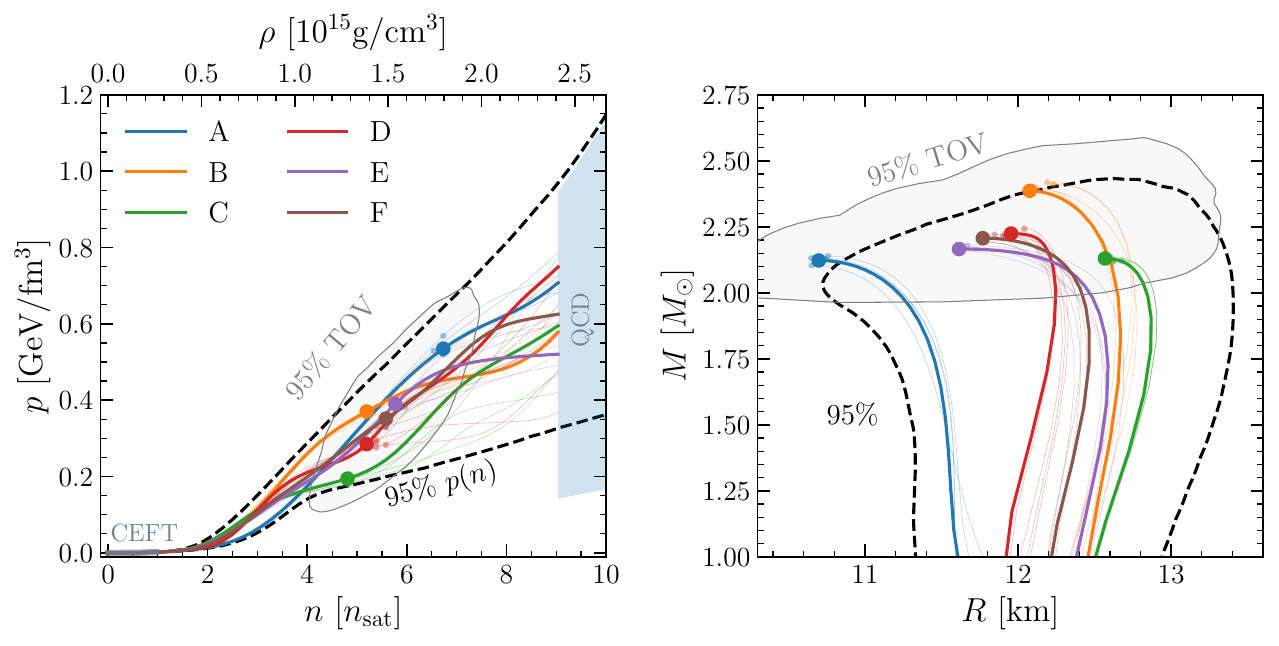}
    \caption{\textbf{Golden and nearby EOSs.} Similar to
      Fig.~\ref{fig:fig1}, with the golden EOSs shown in thick colored
      lines and the TOV points shown with large filled
      points. Additionally shown with thin lines are the five
      next-highest likelihood EOSs from the 30 closest EOSs to the points
      on the $68\%$ contour (see text in SM for more details). The TOV
      points for these nearby EOSs are also shown with small filled
      points.}
\label{fig:pn_mr_nearby}
\end{figure}

Figure~\ref{fig:pn_mr_nearby} shows the golden EOSs selected by this
procedure as curves in the $(p,n)$ and $(M,R)$ planes (thick solid
colored lines), as well as the five next-highest likelihood EOSs from the
30 nearby EOSs (thin solid colored lines). Note that at least for
densities $n \leq n_\text{TOV}$, the selection procedure is robust, with
the nearby EOSs having a similar structure to the corresponding golden
EOSs. Table~\ref{tab0}, instead, provides a concise summary of the most
salient properties of the binaries with the golden EOSs that have been
simulated, together with the characteristic GW frequencies $f_2$ and
$f_{\rm rd}$.

\renewcommand{\arraystretch}{1.2}
\setlength\tabcolsep{3 pt}
\begin{table}
\begin{tabular}{cccccccccccc}
 \toprule
EOS & $M_\mathrm{TOV}$ & $C_{\rm TOV}$ & $R_{1.4}$ & $p_{\rm TOV}$ & $n_{\rm TOV}$ & $q$  & $\tilde\Lambda$ & $f_2$ & ${f}_{\rm rd}$ & ${d{\hat{E}}_\mathrm{GW}}/{d\hat{J}}_\mathrm{GW}$ \\
  & $[M_\odot]$ & & $[{\rm km}]$ &  $[{\rm GeV}/{{\rm fm}^3}]$ & $[n_{\rm
    sat}]$ &  &  & $[{\rm Hz}]$ & $[{\rm Hz}]$ & \\
 \midrule
 A    &  $2.12$  &  $0.293$  &  $11.55$  &  $0.535$  &  $6.73$  &  $0.70$  &  $286$  &  $3380$ &  $3366$  &  $3.48$\\
      &          &           &           &           &          &  $0.85$  &  $301$  &  $3245$ &  $3344$  &  $3.38$\\
      &          &           &           &           &          &  $1.00$  &  $303$  &  $3205$ &  $3229$  &  $3.25$\\[1ex]
 B    &  $2.39$  &  $0.292$  &  $12.57$  &  $0.370$  &  $5.19$  &  $0.70$  &  $528$  &  $2600$ &  $2587$  &  $3.02$\\
      &          &           &           &           &          &  $0.85$  &  $538$  &  $2710$ &  $2601$  &  $2.91$\\
      &          &           &           &           &          &  $1.00$  &  $566$  &  $2725$ &  $2716$  &  $3.00$\\[1ex]
 C    &  $2.13$  &  $0.250$  &  $12.72$  &  $0.195$  &  $4.81$  &  $0.70$  &  $520$  &  $2495$ &  $2471$  &  $2.85$\\
      &          &           &           &           &          &  $0.85$  &  $577$  &  $2615$ &  $2695$  &  $2.85$\\
      &          &           &           &           &          &  $1.00$  &  $627$  &  $2635$ &  $2611$  &  $2.89$\\[1ex]
 D    &  $2.22$  &  $0.275$  &  $12.06$  &  $0.285$  &  $5.19$  &  $0.70$  &  $352$  &  $2680$ &  $2677$  &  $2.93$\\
      &          &           &           &           &          &  $0.85$  &  $358$  &  $2900$ &  $2759$  &  $2.98$\\
      &          &           &           &           &          &  $1.00$  &  $434$  &  $2850$ &  $2883$  &  $3.00$\\[1ex]
 E    &  $2.17$  &  $0.275$  &  $12.53$  &  $0.390$  &  $5.77$  &  $0.70$  &  $520$  &  $2665$ &  $2637$  &  $3.08$\\
      &          &           &           &           &          &  $0.85$  &  $540$  &  $2700$ &  $2687$  &  $2.95$\\
      &          &           &           &           &          &  $1.00$  &  $562$  &  $2756$ &  $2761$  &  $3.05$\\[1ex]
 F    &  $2.21$  &  $0.277$  &  $12.37$  &  $0.352$  &  $5.58$  &  $0.70$  &  $456$  &  $2680$ &  $2662$  &  $2.99$\\
      &          &           &           &           &          &  $0.85$  &  $494$  &  $2735$ &  $2625$  &  $2.94$\\
      &          &           &           &           &          &  $1.00$  &  $502$  &  $2815$ &  $2802$  &  $3.06$\\[1ex]
 DD2  &  $2.41$  &  $0.299$  &  $13.20$  &  $0.544$  &  $5.42$  &  $1.00$  &  $777$  &  $2590$ &  $2574$  &  $2.71$\\[1ex]
 V-QCD &  $2.14$  &  $0.265$  &  $12.47$  &  $0.296$  &  $5.09$  &  $1.00$  &  $565$  &  $2860$ &  $2852$  &  $2.86$\\
 \bottomrule
\end{tabular}
	\caption{\textbf{EOS, NS and BNS properties.} For each EOS, we list the
   TOV-mass $M_{\rm TOV}$, the TOV compactness $C_{\rm TOV}$, the radii
   of a $1.4~M_\odot$ NS $R_{1.4}$, the TOV pressure $p_{\rm TOV}$, the
   TOV number density $n_{\rm TOV}$, the binary tidal deformability
   $\tilde\Lambda$, the post-merger frequencies $f_2$, the long-ringdown
   frequency ${f}_{\rm rd}$ and the corresponding slope
   $d\hat{E}_\mathrm{GW}/d\hat{J}_\mathrm{GW}$.}
 \label{tab0}
\end{table}

\subsection*{Merger Simulations and GW Analysis}

The initial data in our simulations is computed using the spectral-solver
code \texttt{FUKA}~\cite{Papenfort2021b} to generate equal and non-equal
mass irrotational BNS initial data with a separation of $\approx 45\,{\rm
  km}$. \texttt{FUKA} uses the so-called extended conformal thin-sandwich
formulation of Einstein's field equations to solve for binaries in the
quasi-circular orbit approximation. In addition, residual eccentricities
are reduced by applying estimates for the orbital and radial infall
velocities at $3.5$PN order~\cite{Papenfort2021b}.

For the evolution we instead make use of the
\texttt{Einstein-Toolkit}~\cite{EinsteinToolkit_etal:2020_11} that
includes the fixed-mesh box-in-box refinement framework
\texttt{Carpet}~\cite{Schnetter-etal-03b}. More specifically, we use six
refinement levels with the finest grid having a spacing of $295\,{\rm m}$
and impose reflection symmetry across the orbital plane. This
 resolution allows us to explore a reasonable part of the EOS
and BNS parameter space, while keeping the computational costs
affordable. The computational domain has an outer boundary at $\pm
1512\,{\rm km}$, which allows us to compute GWs accurately and impose
suitable boundary conditions.

For the spacetime evolution we use \texttt{Antelope}~\cite{Most2019b},
which solves a constraint damping formulation of the Z4
system~\cite{Alic:2011a}, while we evolve matter using the \texttt{FIL}
general-relativistic magnetohydrodynamic
code~\cite{Most2019b}. \texttt{FIL} implements fourth-order conservative
finite-differencing methods, enabling a precise hydrodynamic evolution
even at the resolution used here. Furthermore, \texttt{FIL}
is able to handle tabulated EOSs that are dependent on temperature and
electron-fraction, and includes a neutrino transport scheme that can
handle neutrino cooling and weak interactions~\cite{Musolino2023}. To
maintain our description as simple as reasonably possible, we have
considered zero magnetic fields and neglected the radiative transfer of
neutrinos; while we do not expect qualitative changes from either
magnetic fields or neutrinos, it is reasonable to expect will play a
quantitative role in establishing the long-ringdown slope.

As discussed in the main text, our agnostic EOS construction provides the
``cold'' (\ie $T=0$, where $T$ is the temperature) part of the EOSs,
while the ``thermal'' part is added during the evolution to account for
shock-heating effects during and after the merger~\cite{Baiotti08}. More
specifically, the total pressure is given by the sum of the cold EOS
$p_{\rm c}=p_{\rm c}(\rho)$ and a thermal component $p_{\rm th}=p_{\rm
  th}(\rho,T)=\rho T$, where $\rho := m_\text{B} n$ is the rest-mass
density and $m_\text{B}=931.5\,{\rm MeV}$ the atomic mass
unit. Analogously, the total specific internal energy can be separated
into a cold $\epsilon_{\rm c}=\epsilon_{\rm c}(\rho)$ and a thermal part
$\epsilon_{\rm th}=\epsilon_{\rm th}(T)$. Here, $\epsilon_{\rm c}(\rho) =
{e_c(\rho)}/{\rho}-1$ and $\epsilon_{\rm th}(T) := {T}/{(\Gamma_{\rm
    th}-1)}$, where $e_c(\rho)$ is the energy density of the cold EOS and
$\Gamma_{\rm th}$ is the thermal adiabatic index, which we choose to take
the constant value $\Gamma_{\rm th}=1.75$. As a result, the total
pressure and total internal energy density are given by
\begin{equation}
p(\rho,T)=p_c(\rho)+\rho T\,,\qquad \qquad
\epsilon(\rho,T)=\frac{e_c(\rho)}{\rho}-1+\frac{T}{\Gamma_{\rm th}-1}\,,
\end{equation}
where the cold contributions $e_c(\rho)$ and $p_c(\rho)$ are provided in
tabulated form by the GP construction explained above. The entropy can
then be expressed as
\begin{equation}
s(\rho,T)=\frac{1}{\Gamma_{\rm th}-1}{\rm ln}
\left(\frac{\bar{\epsilon}_{\rm th}}{\rho^{\Gamma_{\rm th}-1}} \right)\,,
\end{equation}
where $\bar{\epsilon}_{\rm th} := {\rm max}(\epsilon_{\rm th}, s_{\rm
  min})$, with some numeric lower bound for the entropy $s_{\rm
  min}=10^{-10}$. \revtext{In summary, our construction realizes a
  model-independent parametrization for the density and temperature
  dependence of viable NS EOSs without any information on the particle
  composition. Furthermore, since we neglect neutrino emission and
  absorption, no composition dependence is present in our
  simulations. Future analyses with self-consistent temperature-dependent
  EOSs could additionally explore a possible dependence of the
  long-ringdown slope on composition.}

For the GW analysis, we use the Newman-Penrose formalism to relate the
Weyl curvature scalar $\psi_4$ to the second time derivative of the
polarization amplitudes of the GW strain $h_{+,\times}$
via~\cite{Bishop2016}
\begin{equation}\label{eq:psi4}
  \ddot{h}_++i\ddot{h}_\times=\psi_4
  :=\sum_{\ell=2}^{\infty}\sum_{m=-\ell}^{m=\ell}\psi_4^{\ell,m}
     {_{-2}Y}_{\ell,m}\,,
\end{equation}
where $_sY_{\ell,m}(\theta,\phi)$ are spin-weighted spherical harmonics
of weight $s=-2$. From our simulations, we extract the multipoles
$\psi_4^{\ell,m}$ with a sampling rate of $\approx 634\,{\rm kHz}$ from a
spherical surface with radius $\approx 574\,{\rm km}$ centred at the
origin of our computational domain and extrapolate the result to the
estimated luminosity distance of $40\,{\rm Mpc}$ of the GW170817
event~\cite{LIGOScientific:2017vwq}. In addition, we fix the angular
dependence of the spherical harmonics by considering a viewing angle
$\theta=15^{\circ}$, as inferred determined from the jet of
GW170817~\cite{Ghirlanda:2018uyx} and set $\phi=0^{\circ}$ without loss
of generality. We restrict our analysis to the multipoles $\ell \leq 4$
of the expansion~\eqref{eq:psi4} and note that the $\ell=|m|=2$ modes
represent the dominant contribution in our analysis; indeed the relative
difference in the maximum GW amplitude when considering multipoles with
$\ell\leq 4$ and $\ell=2$ is less than $3\%$. Furthermore, we report all
results as functions of the retarded time $t-t_{\rm mer}$, where $t_{\rm
  mer}$ is defined as the time of the global maximum of the GW amplitude
$\sqrt{h_+^2+h_\times^2}$.

An important quantity in our analysis is the instantaneous GW frequency
$f_{\rm GW}$, defined as
\begin{equation}\label{eq:fGW}
    f_{\rm GW}:=\frac{1}{2 \pi} \frac{d \phi}{dt}\,,\qquad \qquad \phi:={\rm
      arctan} \left( \frac{h_\times^{2,2}}{h_+^{2,2}} \right)\,.
\end{equation}
The radiated power is given by the integral expression~\cite{Bishop2016}
\begin{equation}
  \dot{E}_{\rm
    GW}=\frac{r^2}{16\pi}\sum_{\ell=2}^\infty\sum_{m=-\ell}^\ell \left|
  \int_{-\infty}^t dt'\,\psi_4^{\ell m} \right|^2\,,
\end{equation}
where the total emitted GW energy follows from another time integration
$E_{\rm GW}(t) := \int_{-\infty}^{t}dt' \dot{E}_{\rm GW}(t')$. Similarly,
the rate of radiated angular momentum is defined as~\cite{Bishop2016}
\begin{equation}
  \dot{J}_{\rm GW}:=\frac{r^2}{16\pi}{\rm Im}\left\{\sum_{\ell=2}^\infty
        \sum_{m=-\ell}^\ell m \left(\int_{-\infty}^{t}dt'\psi_4^{\ell
          m}\right) \int_{-\infty}^t dt'\int_{-\infty}^{t'}
        dt''\,\bar{\psi}_4^{\ell m}\right\}\,,
\end{equation}
where $r$ is the observer distance and where $\bar{\psi}_4$ is the complex
conjugate of $\psi_4$ and the total emitted angular momentum follows again from
another time integration $J_{\rm GW}(t) := \int_{-\infty}^{t}dt' \dot{J}_{\rm
GW}(t')$.

\begin{figure}
  \centering
  \includegraphics[width=\textwidth]{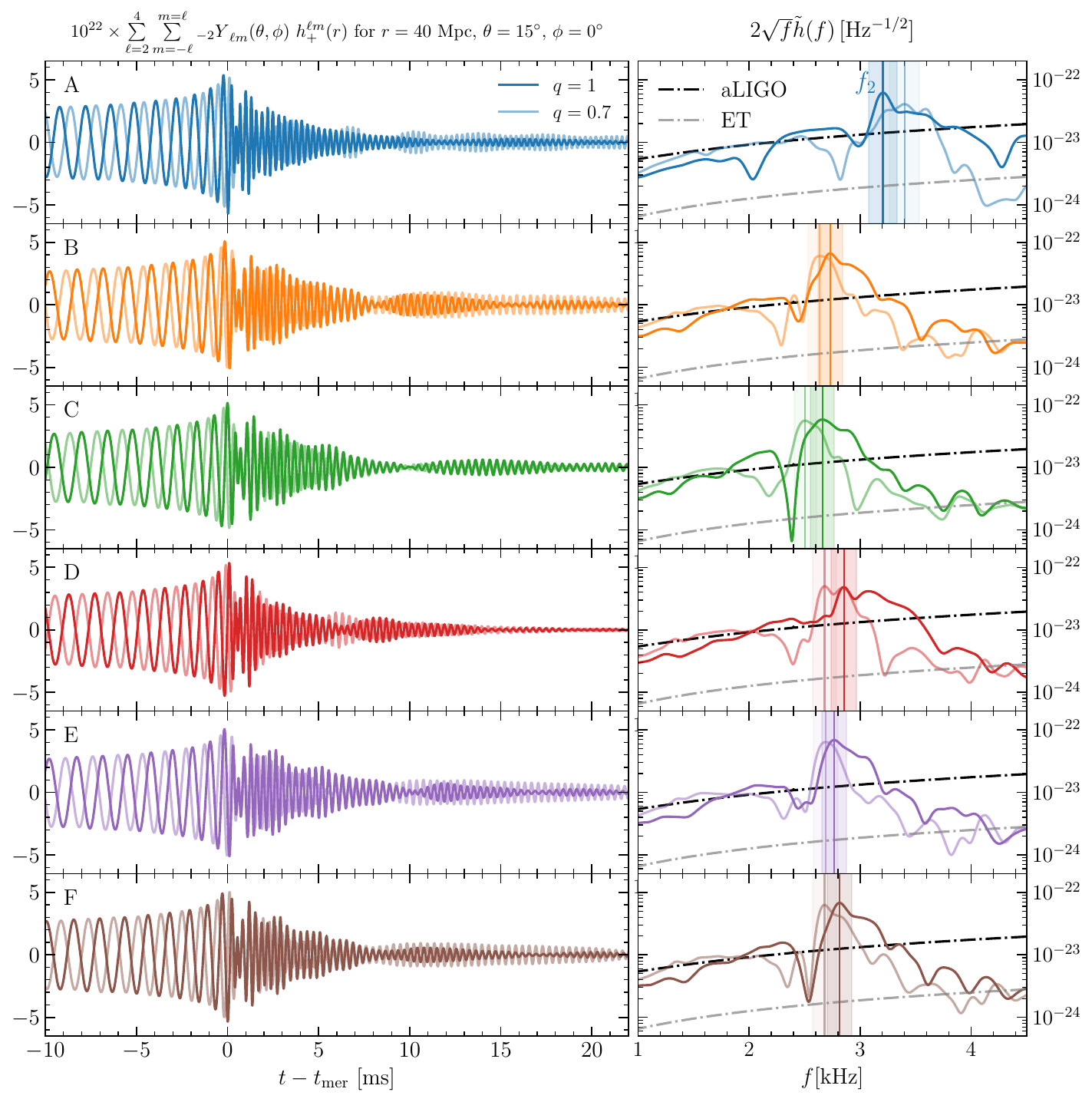}
  \caption{\textbf{Gravitational Waves} GW strain (left) and post-merger
    power spectral density (right) at $40$~Mpc detector distance and
    $\theta=15^{\circ}$ viewing angle for equal chirp mass
    $M_c=1.18\,M_\odot$ and two different mass ratios $q=1,0.7$. {In the
      right panel we mark with solid lines the dominant post-merger
      frequency $f_2$}, where the shaded areas indicate a
      $8\%$ relative error estimate.}
\label{fig:GW}
\end{figure}

In the main text we work with dimensionless energies $E_{\rm
  GW}(t)/E_{\rm GW}^{\rm mer}$ and angular momenta $J_{\rm GW}(t)/J_{\rm
  GW}^{\rm mer}$ obtained by normalizing with the respective values at
merger time $E_{\rm GW}^{\rm mer}:=E_{\rm GW}(t_{\rm mer})$ and $J_{\rm
  GW}^{\rm mer}:=J_{\rm GW}(t_{\rm mer})$. When expressed in terms of
strain components, and in full generality, the ratio of the radiated
energy and angular-momentum rates ${dE_{\rm GW}}/{dJ_{\rm GW}}$ is
similar to the instantaneous GW frequency
\begin{equation}
  \label{eq:fGWvsdEJ}
  \frac{dE_{\rm GW}}{dJ_{\rm GW}}=\frac{\dot{E}_{\rm GW}}{\dot{J}_{\rm
      GW}}=\frac{\dot{h}_+^2+\dot{h}_\times^2}{h_+\dot{h}_\times-\dot{h}_+h_\times}\,,
  \qquad \qquad
  f_{\rm GW}=\frac{1}{2\pi}
    \frac{h_+\dot{h}_\times-\dot{h}_+h_\times}{h_+^2+h_\times^2}\,.
\end{equation}
For a simple system with an $\ell=2, m=2$ deformation, \eg a compact
rotating system with eccentric mass distribution like the toy model of
Ref.~\cite{Takami2015} and for which $h_+(t)\propto \cos(\phi(t))$ and
$h_\times(t)\propto \sin(\phi(t))$ with GW phase $\phi(t)$, one obtains
the identity ${\dot{E}_{\rm GW}}/{\dot{J}_{\rm GW}}=f_{\rm GW} / (2
\pi)$. Since in the long ringdown $f_{\rm GW}(t) \simeq {\rm const.} =:
f_{\rm rd}$, expressions~\eqref{eq:fGWvsdEJ} explain why the radiated
energy and angular momentum are linearly related.

Finally, we analyze the spectral features of the waveforms and compute
the power spectral density (PSD) of the signal as~\cite{Takami2015}
\begin{equation}
  \label{eq:PSD}
  \tilde{h}^{\ell,m}(f):=\frac{1}{\sqrt{2}}\biggl(
  \left|\int\! dt\,{\rm e}^{-2\pi ift} h^{\ell,m}_+(t)\right|^2 +
  \left|\int\! dt\,{\rm e}^{-2\pi ift} h^{\ell,m}_\times(t)\right|^2
  \biggr)^{1/2}\,,
\end{equation}
where the time integration is performed over the interval $t-t_{\rm
  mer}\in [0, 30]\,{\rm ms}$ or up to the time at which a black hole is
formed if the post-merger remnant collapses earlier. \revtextB{As done
  routinely~(see, \eg \cite{Bauswein2015, Takami2015, Rezzolla2016,
    DePietri2018, Kiuchi2022, Breschi2022a}), the dominant post-merger
  frequency $f_2$ is then determined by the global maximum of the PSD.}

Figure~\ref{fig:GW} summarises the GW output from a number of our
simulations by reporting on the left column the GW strain and on the
right column the corresponding PSD from the post-merger signal when
compared with the estimated sensitivities of advanced LIGO (aLIGO) and
the Einstein Telescope (ET). The data refers to BNS simulations for the
EOSs ${\rm A}$-${\rm F}$ (top to bottom), all having the same chirp mass
$\mathcal{M}_c = 1.18\,M_\odot$ and two different mass ratios $q=1,0.7$
(dark and light colors, respectively). Consistent with the expectations
from the GW170817 event, the results shown assume a distance of $40\rm
Mpc$ and a viewing angle of $\theta=15^\circ$.

\begin{figure}
  \centering
  \includegraphics[width=0.5\textwidth]{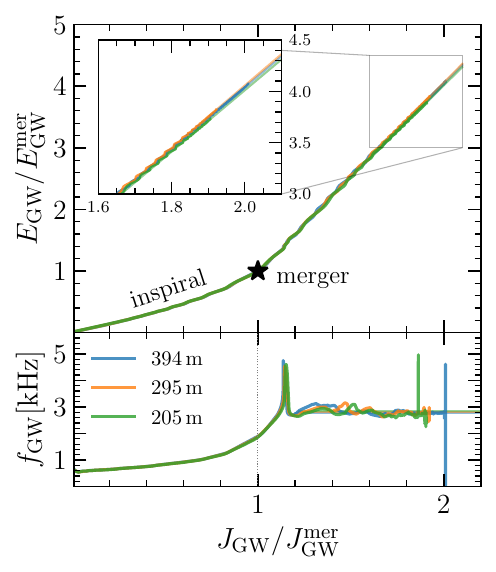}
   \caption{\textbf{Resolution dependence.} GW energy as function of the
     angular momentum (top) and the corresponding GW frequency (bottom)
     for different values of the grid resolution. The examples shown are
     for equal-mass binaries and the EOS F.}
\label{fig:convergence}
\end{figure}

\revtext{Next, we demonstrate the robustness of our long-ringdown slope
  computation with respect to the grid resolution used in the numeric
  simulation. To this scope, we performed, in addition to our standard
  resolution ($295\,{\rm m}$), also simulations with higher ($205\,{\rm
    m}$) and lower ($394\,{\rm m}$) resolutions. The results of these
  simulations are summarized in Fig.~\ref{fig:convergence}, which
  highlights how the slope is essentially insensitive 
  to the resolution and that even simulations with low resolutions result in
  slope values that are well within the uncertainty of the fit ($\pm
  0.1$, see discussion below). More specifically, we measure slopes of
  $\{2.99\pm0.06, 3.06\pm 0.07, 3.02\pm0.06 \}$ for grid-resolutions of
  $\{205\,{\rm m}$, $295\,{\rm m}$, $394\,{\rm m}\}$, respectively. More
  importantly, the measure of the slope at different resolutions displays
  a much smaller variance than the equivalent measure of $f_{\rm rd}$
  (see lower panel of Fig.~\ref{fig:convergence}). The somewhat
  surprising robustness of the long-ringdown slope with resolution can be
  simply explained by the fact that the post-merger waveform is dominated
  by the large-scale $\ell=2, m=2$ deformations of the merger remnant,
  which are only weakly influenced by the small-scale features within the
  remnant.}

\revtext{Our final discussion on the methods employed in our analysis is
  focussed on the accuracy of the slope extraction from the numerical
  simulations. This operation requires first to identify the optimal
  time-range for the linear least-squares fit of $E_{\rm GW}(J_{\rm
    GW}(t))$, whose starting and final times are determined by minimizing
  the variance of the linear fit. More specifically, we first compute the
  slope and its variance for a number of different starting times $t_{\rm
    in} - t_{\rm mer} \in [1-10]\,{\rm ms}$ using a fixed value for the
  final time $t_{\rm fin} - t_{\rm mer} = 15\,{\rm ms}$. In this way, we
  found that a starting time of $t_{\rm in} - t_{\rm mer} = 1\,{\rm ms}$
  results in an approximate variance of $\pm 0.1$ for the slope. Larger
  values for the starting time, \eg $5\,{\rm ms}$ or $10\,{\rm ms}$,
  result in significantly larger variances of $\pm 0.7$ and $\pm 3.1$,
  respectively. Next, we compute the slope and its variance for various
  values of the final time $t_{\rm fin} - t_{\rm mer} \in [2-25]\,{\rm
    ms}$ while keeping the starting time fixed at $t_{\rm in} - t_{\rm
    mer} = 1\,{\rm ms}$. In this way, we found that the variance
  saturates at $t_{\rm fin} - t_{\rm mer} \approx 15\,{\rm ms}$ to values
  similar to those obtained by varying the starting time. Increasing the
  final time of the fit does not lead to further improvement of the fit
  quality and this is because it becomes increasingly difficult to
  accurately compute the small changes in $E_{\rm GW}$ and $J_{\rm GW}$
  at times beyond $t - t_{\rm mer} \gtrsim 15\,{\rm ms}$, where the
  amplitude of the GW signal becomes very small.}
  \revtextB{We apply an analogous procedure to determine $f_{\rm rd}$.}

\subsection*{Performing the mock measurement}

We perform a mock joint measurement of $f_2$ and the slope $d
\hat{E}_\mathrm{GW} / d \hat{J}_\mathrm{GW}$ in the following manner. We
assume a measurement whose uncertainty we model with a multivariate
Gaussian distribution of $f_2$ and $d \hat{E}_\mathrm{GW} / d
\hat{J}_\mathrm{GW}$ for simplicity, as well as a uniform measurement of
$q \in [0.7, 1.0]$. Let us denote the joint likelihood from the
measurement $P_\mathrm{meas}(\mathrm{data} | f_2, d \hat{E}_\mathrm{GW} /
d \hat{J}_\mathrm{GW}, q)$. First, we fit a two-component model to the
$f_2$ and $d \hat{E}_\mathrm{GW} / d \hat{J}_\mathrm{GW}$ data, the
posterior of which we denote by  $P_\mathrm{mod}(d \hat{E}_\mathrm{GW} /
d \hat{J}_\mathrm{GW} | \mathrm{EOS}, q)$ respectively. This
two-component model is just the product of two models of the form
\eqref{eq:bilinear_model} for $f_2$ and $d \hat{E}_\mathrm{GW} / d
\hat{J}_\mathrm{GW}$. Next, we compute the likelihood that each EOS is
consistent with the mock measurement by evaluating 
\begin{align}
    P(\mathrm{data} | \mathrm{EOS}, q) =  \int & 
    d f_2 \, d\! \left(d \hat{E}_\mathrm{GW} / d \hat{J}_\mathrm{GW}\right) \nonumber \\
    &\times P_\mathrm{meas}(\mathrm{data} | f_2, d \hat{E}_\mathrm{GW} / d \hat{J}_\mathrm{GW}, q) \nonumber \\
    &\times P_\mathrm{mod}(d \hat{E}_\mathrm{GW} / d \hat{J}_\mathrm{GW}, f_2 | \mathrm{EOS}, q)\,,
    \label{eq:joint_meas}
\end{align}
by Monte-Carlo sampling of the measurement distribution, which we then
use in Bayes's theorem to generate the posteriors in
Fig.~\ref{fig:pn_mr_measured}, using a flat prior on $q \in [0.7, 1.0]$.
The likelihoods when using only one of the two bilinear models is defined
similarly.

\subsection*{On the robustness of the correlation}

In order to identify potential degeneracies in the long-ringdown slope
between the chirp mass and the EOS properties we simulate the equal-mass
($q=1$) binaries with EOS ${\rm A}$ and ${\rm C}$ with three different
values for the chirp mass, namely, $\mathcal{M}_{\rm chirp} =
1.13,\,1.18,\,1.22\,M_\odot$. Since the chirp mass represents one of the
best-measured quantities in BNS mergers with a few-percent error, what we
are assessing in this way is the dependence for a given EOS of the
long-ringdown slope on $\mathcal{M}_{\rm chirp}$. Stated differently, we
can assess how different long-ringdown slopes cluster when exploring the
possible ranges in the chirp mass.

\begin{figure}
  \centering
  \includegraphics[width=0.6\textwidth]{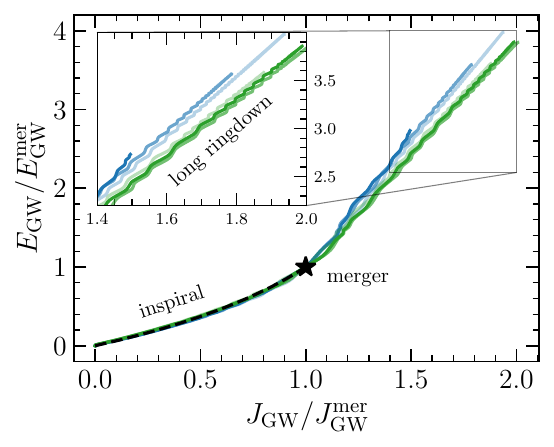}
  \caption{\textbf{Impact of chirp mass} relation between normalized GW
    energy and angular momentum for model A (blue) and C (green) for
    three different values of the chirp mass $\mathcal{M}_{\rm chirp} =
    1.13, 1.18, 1.22\,M_\odot$ (light to dark colors) and fixed mass
    ratio $q=1$. }
\label{fig:GWMc}
\end{figure}

The result of this test are displayed in Fig.~\ref{fig:GWMc}, which is
similar to Fig.~\ref{fig:fig2}, but where we show the slope for model
${\rm A}$ in green and for model ${\rm C}$ in blue, while light to dark
colors indicate small to large values of $\mathcal{M}_{\rm chirp}$,
respectively. Evidently, the variations in the chirp mass lead to
significantly smaller differences in the long-ringdown slope than those
introduced by the EOSs. Hence, Fig.~\ref{fig:GWMc} highlights that the EOS
represents the dominant contribution to the long-ringdown slope and that
the chirp mass plays only a sub-dominant role. This is natural to expect
since the long-ringdown slope is essentially set by the equilibrium of
the HMNS, which, in turn, is predominantly determined by the EOS.

\begin{figure}
  \centering
  \includegraphics[width=0.97\textwidth]{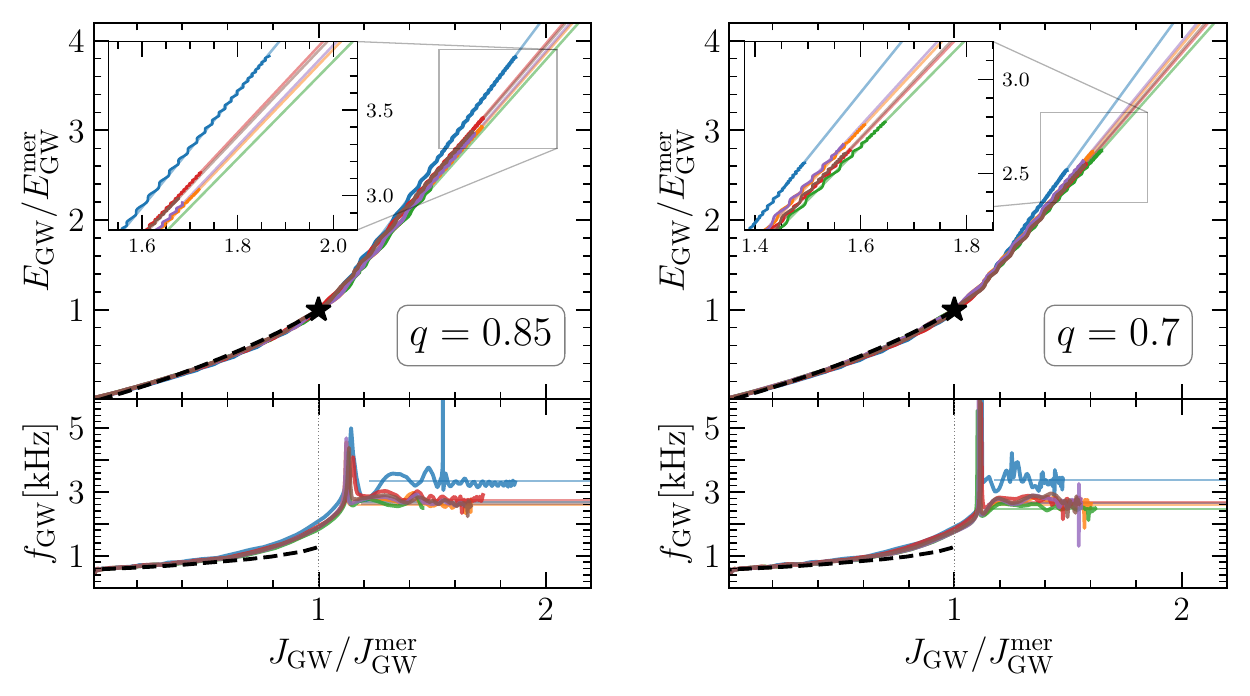}
    \caption{\textbf{Unequal-mass binaries} Evolution of the normalized
      GW energy and GW frequency as a function of the normalized radiated
      angular momentum for the case of asymmetric binaries with mass
      ratio $q = 0.85$ (left) and $q = 0.7$ (right). }
\label{fig:fig2_qneq1}
\end{figure}

\revtext{Next, we show in Fig.~\ref{fig:fig2_qneq1} results analogue to
  Fig.~\ref{fig:fig2}, but for mass ratios $q=0.85$ and $q=0.7$, which
  complement the information shown in Fig.~\ref{fig:fig3}. Note that also
  the unequal-mass binaries show a clear linear correlation in the
  radiated energy and angular momentum during the long ringdown and that
  different mass ratios lead to slightly different slopes.}  We should
also remark that over the timescale considered here, the $\ell=2, m=2$
mode is still the dominant one and the contributions from the $\ell=2,
m=1$ mode are at least two orders of magnitude
smaller. \revtextB{However, it is possible at later times that the
  $\ell=2, m=1$ mode will dominate (see, \eg~\cite{Topolski2024}) both
  for highly asymmetric binaries (for which the $m=1$ deformation is
  quite large right after merger), but also for equal-mass binaries (for
  which the $m=1$ asymmetry is initially small but grows steadily). Also
  for this $\ell=2, m=1$ GW mode, the radiated energy and angular
  momentum will remain linearly related, albeit with a different
  (smaller) slope.}

\begin{figure}
  \centering
  \includegraphics[width=0.6\textwidth]{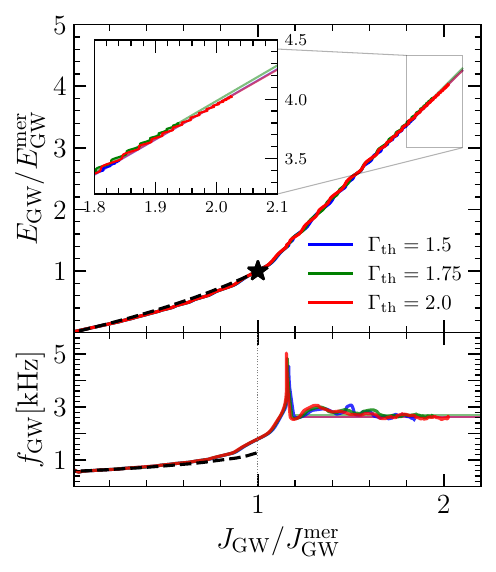}
  \caption{\textbf{Impact of $\Gamma_{\rm th}$} Same as
    Fig.~\ref{fig:GWMc} but for model B with three different values of
    $\Gamma_{\rm th}=1.5,1.75,2.0$ and fixed binary parameters
    $\mathcal{M}_{\rm c}=1.18$ and $q=1$.}
\label{fig:GWGamma}
\end{figure}

Another potential source of uncertainty in our results may have come from
the thermal part of the EOS, which is admittedly simplified but
qualitatively correct. In order to assess the impact of the thermal
contributions, we study a reference EOS $B$ with three different values
that span the possible range expected for the adiabatic index, \ie
$\Gamma_{\rm th}=1.5,1.75,2.0$. The results of this analysis are reported
in Figure~\ref{fig:GWGamma}, which shows how larger values of
$\Gamma_{\rm th}$ typically lead to higher thermal-pressure
contributions, which help support the $\ell=2=m$ deformation of the
merger remnant, and thus result in a more efficient GW emission in the
post-merger phase. At the same time, the long-ringdown slope is
essentially unaffected by the choice of $\Gamma_{\rm th}$ as can be seen
from the tight overlap of the corresponding curves. As a result, we can
conclude that our choice of a fiducial value of $\Gamma_{\rm th}=1.75$
does not introduce any bias on the reported long-ringdown slopes.

\begin{figure}[htb]
  \centering
  \includegraphics[width=0.6\textwidth]{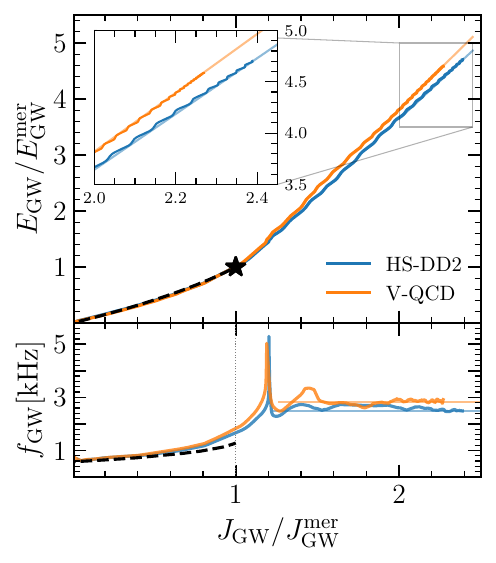}
  \caption{\textbf{Temperature dependent EOSs} Same as
    Fig.~\ref{fig:GWMc}, but for two models with tabulated temperature
    dependence, namely the HS-DD2 EOS and the intermediate variant of the
    holographic V-QCD EOS. Both simulations assume $q=1$ and
    $\mathcal{M}_{\rm chirp}=1.18~M_\odot$. }
\label{fig:EOS_Tdep}
\end{figure}

Finally, in Fig.~\ref{fig:EOS_Tdep} we show results for two EOSs with a
tabulated temperature dependence, namely the Hempel-Schaffner DD2
(HS-DD2) EOS~\cite{Hempel2010} and the intermediate variant of the
holographic Veneziano QCD (V-QCD) EOS~\cite{Demircik:2021zll}. While the
HS-DD2 EOS models purely hadronic matter, the V-QCD EOS features a strong
first-order phase transition from hadronic to quark matter that induces
the HMNS to collapse to a black hole in this simulation. As shown
in~\cite{Tootle2022}, the strong phase transition of the V-QCD EOS does
not allow stable quark-matter cores inside isolated stars, but a
significant amount of quark matter can be formed during the metastable
post-merger phase. Importantly, neither the presence of quark matter, nor
the microscopic prescription for the temperature dependence in these EOSs
alter the basic features of the long ringdown.

\begin{table}
\begin{tabular}{rrrrrrrrr}
    \toprule
        &$\beta_0$& $\beta_1$& $\beta_2$& $\beta_3$& $\beta_4$& $\beta_5$& $\beta_6$& \\
    \midrule
        $\beta_0$&$  4.16$&$       $&$       $&$       $&$       $&$       $&$       $ \\
        $\beta_1$&$ -6.40$&$  10.59$&$       $&$       $&$       $&$       $&$       $ \\
        $\beta_2$&$  2.57$&$  -6.32$&$  11.27$&$       $&$       $&$       $&$       $ \\
        $\beta_3$&$ -3.86$&$   6.24$&$  -4.52$&$   4.48$&$       $&$       $&$       $ \\
        $\beta_4$&$  6.15$&$ -10.68$&$   9.15$&$  -7.22$&$  12.46$&$       $&$       $ \\
        $\beta_5$&$ -4.39$&$   8.85$&$  -9.86$&$   5.29$&$ -10.48$&$  10.97$&$       $ \\
        $\beta_5$&$  1.76$&$  -2.08$&$  -3.25$&$  -0.07$&$  -0.08$&$   0.41$&$   3.83$ \\
    \bottomrule
\end{tabular}
    \caption{\textbf{Covariance matrix of the bilinear model} The
      covariance matrix $\operatorname{cov}(\bm{\beta})$ of the trained
      bilinear model used in this work in
      Eq.~\protect\eqref{eq:bilinear_model}. The covariance matrix is
      symmetric. Here, the model has been trained with inputs where
      $p_\mathrm{TOV}$ is in units of ${\rm GeV/fm}^{3}$ and
      $n_\mathrm{TOV}$ is in ${\rm fm^{-3}}$.}
    \label{tab1}
\end{table}

\bibliography{aeireferences.bib}

\end{document}